\documentclass[twocolumn,showpacs,preprintnumbers,amsmath,amssymb]{revtex4}

\newcommand{\commentoutA}[1]{}

\usepackage{graphicx}
\usepackage{dcolumn}
\usepackage{bm}
\begin{document}

\preprint{LA-UR 04-5100}

\title{Non-Orthogonal Density Matrix Perturbation Theory}

\author{Anders M. N. Niklasson$^{(1),(\dagger)}$} 
\author{Val\'ery Weber$^{(1),(2)}$, and Matt Challacombe$^{(1)}$}
\affiliation{$^{(1)}$Theoretical Division, Los Alamos National Laboratory, Los Alamos, New Mexico 87545}
\affiliation{$^{(2)}$Department of Chemistry, University of Fribourg, 1700 Fribourg, Switzerland}

\date{\today}

\begin{abstract}
Density matrix perturbation theory [Phys.\ Rev.\ Lett.  {\bf 92}, 193001 (2004)]
provides an efficient framework for the linear scaling computation of response properties
[Phys.\ Rev.\ Lett.  {\bf 92}, 193002 (2004)].  In this article, we generalize density matrix 
perturbation theory to include properties computed with a perturbation dependent non-orthogonal 
basis.  Such properties include analytic derivatives of the energy with 
respect to nuclear displacement, as well as magnetic 
response computed with a field dependent basis.  The non-orthogonal density matrix 
perturbation theory is developed in the context of recursive purification methods, which are 
briefly reviewed.
\end{abstract}

\pacs{02.70.-c, 31.15.-p, 71.15.-m, 71.15.Dx}
\keywords{electronic structure theory, response, perturbation theory, density
matrix derivative, density matrix, linear scaling electronic structure theory, 
purification, sign matrix, O(N), basis set, orthogonal, non-orthogonal, transformation, 
matrix inverse, Green's function, recursion, commutation relations, Fermi operator}
\maketitle

\section{Introduction}

During the last decade a new computational paradigm has evolved in electronic structure
theory, where no critical part of a calculation is allowed to increase in complexity more than linearly 
with system size 
\cite{WYang92,GGalli92,FMauri93,POrdejon93,XLi93,EStechel94,SGoedecker94,RSilver94,LWang94,JKim95,YWang95,IAbrikosov96,GGalli96,WKohn96,DBowler97,DSanchezportal97,ESchwegler97,MChallacombe97,SYokojima98,RBaer98,CGuerra98,APalser98,SGoedecker99,EArtacho99,GScuseria99,POrdejon00,SWu02,ANiklasson02A,CYam03,ANiklasson04,VWeber04,MWatson04,CTymczak04A,CTymczak04B}. 
Linear scaling electronic structure theory extends tight-binding, Hartree-Fock, and Kohn-Sham schemes
to the study of large complex systems beyond the reach of conventional methods.  
In general, conventional methods have three computational bottlenecks:
(1) construction of the Hartree-Fock, tight-binding, or Kohn-Sham Hamiltonian, 
(2) solution of the self-consistent-field equations to obtain the  ground state, and quite often,
(3) the evaluation of response properties.
Here we will focus on the last problem, including calculations of both linear
and higher order non-linear response. The main purpose is to present
a non-orthogonal generalization of the density matrix perturbation theory,
recently introduced by the authors \cite{ANiklasson04}, which was used for linear 
scaling computation of static electric polarizabilities by perturbed projection 
\cite{VWeber04}.  

A non-orthogonal generalization of the density matrix perturbation theory is important because 
it includes basis set dependent perturbations in the overlap matrix, i.e. the basis function 
inner products, when using non-orthogonal local orbitals.
In this case, the perturbed  Hartree-Fock or Kohn-Sham eigenvalue problem may be expressed 
in the generalized form
\begin{equation}\label{SEqP}
\left( {\rm H} + {\rm H'}\right) \phi_i = \varepsilon_i \left( {{\rm S}+{\rm S}'}\right) \phi_i,
\end{equation}
with both the effective Hamiltonian ${\rm H}$ and the overlap matrix ${\rm S}$
modified by the perturbations ${\rm H'}$ and ${{\rm S}'}$, respectively.
Such perturbations are encountered when using a basis of atom centered Atomic Orbitals and
computing geometric energy derivatives \cite{PPulay69,RAmos89}, and also when calculating magnetic
response properties with field dependent Gauge Including Atomic Orbitals.

Previously, non-orthogonal generalizations of linear scaling methods for calculation of the 
density matrix have been developed by including the overlap matrix as a metric tensor in operator products 
\cite{RNunes94,CWhite97,APalser98}.  Non-orthogonal density matrix schemes for solving the 
coupled-perturbed self-consistent-field equations \cite{COchsenfeld97,HLarsen01,COchsenfeld04} 
have also been put forward, which pose density matrix derivatives implicitly through a set of 
commuting Sylvester-like equations \cite{JBrandts01}.  In this article,  we present a non-orthogonal 
generalization of density matrix perturbation theory, based on an explicit recursive expansion of
the non-orthogonal density matrix and its derivatives.  This theory provides a framework for extending 
linear scaling response calculations to properties with a non-orthogonal basis set dependence on 
the perturbation.

The article is outlined as follows: First we give a brief review of the orthogonal 
formulation of density matrix purification and density matrix perturbation theory. 
Operators represented in an orthogonal basis set are described by italics ($P$). 
Thereafter we generalize the description to non-orthogonal representations, where 
the matrices are described by normal letters (P). The central result is the non-orthogonal
density matrix perturbation theory in section \ref{NOPRT}.
A simple example is given in detail for the expansion of the interatomic pair 
interaction of the diatomic H$_2^+$ molecule up to fourth order.

\section{Orthogonal Density Matrix Purification and Density Matrix Perturbation Theory}

\subsection{Orthogonal purification}

Linear scaling electronic structure theory is based on the quantum locality (or nearsightedness) of 
non-metallic systems \cite{WKohn96,WKohn59,RBaer97,IBeigi99,MHastings04}.
In a local basis, this locality is manifested in an approximate 
exponential decay of density matrix elements with inter-atomic separation.  Under these 
conditions, and using a sparse linear algebra with the dropping of numerically small elements 
below a threshold $\tau$, the number of non-zero entries in the density matrix scales asymptotically
linearly, $O(N)$, with system size $N$.  This sparsity is used to achieve an $O(N)$ complexity in
an iterative construction of the density matrix by operating only with sparse intermediate matrices.
Several of these techniques are based on the Fermi operator relation between the density matrix 
$P$ and the Hamiltonian $H$ taken at $T=0$,
\begin{equation}\label{DM}
P = \theta(\mu I -  H),
\end{equation}
given by the step function $\theta$ (spectral projector), with the step formed at the chemical potential $\mu$.
The chemical potential determines the occupied states via Aufbau filling.   

One approach to constructing $P$ is through expansion of $\theta$ 
using the Chebychev polynomials $T_k$
\cite{RSilver96,AVoter96,WLiang03};
\begin{equation}\label{Cheb}
P = \theta(\mu I -H) \approx \sum_{k=0}^p a_kT_k(H).
\end{equation}
With the two-term recurrence 
\begin{equation}
T_{k+1}(X) = 2XT_k(X)-T_{k-1}(X),
\end{equation}
the computational cost for evaluation of Eq.~(\ref{Cheb}) scales as $O(p)$ with
polynomial order $p$ of the Chebychev expansion. However, this cost can be reduced to 
$O({\sqrt p})$ by using a hierarchical summation of polynomial terms, rather than the 
two-term recurrence relation \cite{MPaterson73,WLiang03}. While this reduces the 
computational cost, it demands more intermediate memory to store temporary matrices.  
In either case, the order of the polynomial approximation in Eq.~(\ref{Cheb}), $p$, must 
be kept fairly low, leading to problems with incompleteness.  An incomplete 
Chebychev series results in Gibbs oscillations, which are high frequency ripples 
that form about the step. These oscillations can be reduced by applying Gibbs damping 
factors to the Chebychev polynomials as in the Kernel Polynomial Method 
\cite{RSilver94,RSilver96,AVoter96}. However, this reduces the slope of the
step function and a higher order expansion is necessary to resolve the step at
the chemical potential \cite{RSilver96}.
 
Alternatively, density matrix purification is a recursive approach to spectral projection
\cite{APalser98,ANiklasson02A,ANiklasson04,ANiklasson03,ANiklasson03B,RMcWeeny60,WClinton69,AHolas01,DMazziotti03}
that approximates the matrix step function projection as
\begin{equation}\label{Rec}
P = \theta(\mu I -H) = \lim_{n \rightarrow \infty} F_n(F_{n-1}(...F_0(H)...)).
\end{equation}
Initiating a purification sequence is the linear transform $F_0(H)$, which  normalizes
the spectra of $H$ to $[0,1]$ in reverse order. The functions $F_n$ ($n>0$) are typically low order,
monotonically increasing polynomials, with fixed points at $0$ and $1$. Each purification
polynomial $F_n$ gradually shifts the eigenvalues of the approximate intermediate 
density matrix  $X_{n}$ to $0$ for unoccupied states and to $1$ for occupied states as 
\begin{equation}
X_{n+1} = F_{n}(X_{n})= \ldots = F_{n}(...F_0(H)...) \, ,
\end{equation}
creating a successively more ``purified'' intermediate $X_{n+1}$. 

\begin{figure}[t]
\resizebox*{3.0in}{!}{\includegraphics[angle=00]{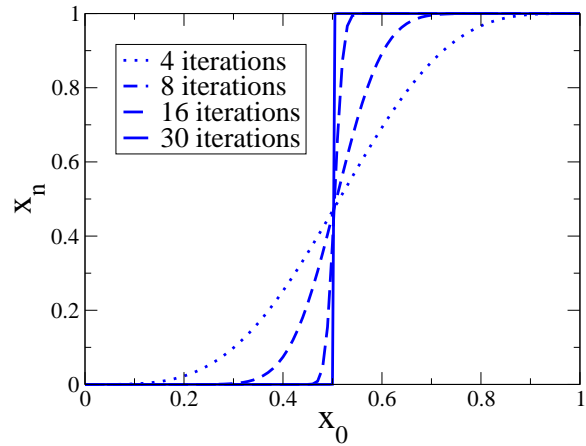}}
\caption{\label{Step_TC2}
The iterative expansion of the step function using second order trace correcting 
purification, Eq.\ (\ref{TC2}).}
\end{figure}

Purification has several advantages.  First, 
a $p^{\rm th}$ order truncated purification sequence can be developed with a complexity of $O(\log p)$.   
For example, with quadratic polynomials,  a $p \sim 10^9$ order expansion can be reached 
in only 30 iterations.  Also, because the polynomials $F_n$ are  monotonically increasing, so to are 
the corresponding purification sequences, regardless of the degree of incompleteness.  The Gibbs oscillations 
resulting from truncation of the Chebychev series are therefore avoided and 
the application of damping factors is no longer necessary. Figure \ref{Step_TC2} illustrates the typical 
behavior using second order trace correcting purification as described below.
                                                                                               
To achieve a linear scaling with purification, thresholding is applied after each recursive 
expansion step to the intermediates $X_n$, removing elements below a tolerance $\tau$ ($\sim 10^{-4}-10^{-6}$). 
This often leads to a substantial increase in computational efficiency, but also
to an accumulation of numerical error with each recursive purification step.
This error involves corruption of the eigenbasis, which at first increases exponentially \cite{ANiklasson03}. 
However, this error accumulation disappears as the eigenvalues of $X_n$ approach 
$0$ or $1$. Since the number of purification steps necessary to reach convergence scales with the 
logarithm of the inverse band gap, the method is stable and the total accumulated error is well controlled.
At convergence the density matrix error scales linearly with the threshold $\tau$
and the error in total energy decreases quadratically with decreasing $\tau$ \cite{ANiklasson03}.

Density matrix purification methods differ in the way the purification polynomials $F_n(X_n)$
are chosen. In grand canonical schemes \cite{APalser98,AHolas01,ANiklasson02A} the initial linear normalization
$X_1 = F_0(H)$ shifts the  eigenvalues such that all occupied eigenvalues are in $[c,1]$ and
all unoccupied eigenvalues in [$0,c]$, where $c$ is some predefined number (typically $c = 0.5$).
Thereafter a fixed purification polynomial with inflection point at $c$ is used, which shifts
eigenvalues above (below) $c$ to $1$ ($0$). At convergence the correct occupation is therefore reached, with
\begin{equation}
{\rm Tr}(P) = \lim_{n \rightarrow \infty} {\rm Tr}(X_n) = N_{\rm e}.
\end{equation}
The problem with this approach is that it requires prior knowledge of the 
chemical potential $\mu$, which has to be shifted to the inflection point $c$ in the initial
normalization $X_1 = F_0(H)$.
To avoid the problem with an unknown chemical potential Palser and Manolopoulos (PM) devised  a
canonical purification scheme \cite{APalser98} with the purification polynomials chosen
such that the trace, i.e. the occupation, is preserved in each purification step.
By choosing the initial normalization such that ${\rm Tr}(X_1) = {\rm Tr}[F_0(H)] = N_{\rm e}$ the PM scheme
automatically converges to the correct density matrix, without prior knowledge of 
the chemical potential.  The problem with this method is that it has a very slow
convergence  at high or low occupation \cite{APalser98,ANiklasson02A}.
A solution to this problem was given by the introduction of trace correcting purification 
\cite{ANiklasson02A} described below.

\subsubsection{Second order trace correcting purification}

Trace correcting purification \cite{ANiklasson02A,ANiklasson03,DMazziotti03} is an 
efficient approach to density matrix purification at both high and low occupation and 
does not require knowledge of the chemical potential. In trace correcting 
purification the polynomials $F_n(X_n)$ correct the trace ${\rm Tr}(X_n)$ and
expand the step function simultaneously.  At convergence, the correct occupation of the 
density matrix is reached, such that ${\rm Tr}(P) = \lim_{n \rightarrow \infty} {\rm Tr}(X_n) = N_{\rm e}$.
The simplest and most memory efficient form is the second order trace correcting algorithm \cite{ANiklasson02A},
given by
\begin{equation} \label{Pract1}
X_1 = F_0(H) = \frac{\varepsilon_{\rm max}I-H}{\varepsilon_{\rm max} - \varepsilon_{\rm min}},
\end{equation}
\begin{equation}\label{TC2}
X_{n+1} = F_n(X_n) = X_n + \sigma_n (I-X_n)X_n,
\end{equation}
where
\begin{equation}\label{Pract2}
\sigma_n = {\rm sign}(N_{\rm e } - N_{\rm occ}),
\end{equation}
\begin{equation}\label{Occupation}
N_{\rm occ}={\rm Tr}(X_n),
\end{equation}
and
\begin{equation}
P = \lim_{n \rightarrow \infty} X_n.
\end{equation}
The constants $\varepsilon_{\rm max}$ and $\varepsilon_{\rm min}$ are upper and lower estimates of the spectral
bounds of $H$, given for example by Gersgorin bounds \cite{APalser98}. The initial
normalization $F_0(H)$ thus transforms all eigenvalues of $H$ to the interval $[0,1]$ in 
reverse order.  The function ${\rm sign}(x)$ denotes the sign of $x$. It is $+1$ if
$x>0$, otherwise it is $-1$.  The trace correcting recursion in Eq.~(\ref{TC2}) is
equivalent to previously published versions \cite{ANiklasson02A,ANiklasson04,VWeber04},
but is presented here in a form more closely related to an efficient implementation.

In the case of aggressive thresholding and high or low occupation, the  eigenvalues of $X_n$
can sometimes be pushed out of the domain of guaranteed convergence, $[0,1]$. 
To enhance stability under loose thresholding, we alternate the sign of the trace correction in every step, 
with $\sigma_n = -\sigma_{n-1}$ as convergence is approached, typically when ${\rm Tr}[(I-X_n)X_n] < 0.1$.

\subsection{Orthogonal Perturbation Theory}

The main obstacle in formulating a density matrix perturbation theory 
based directly on the  relation between the Hamiltonian and the spectral projector,
given  by Eq.~(\ref{DM}), is the discontinuous, non-analytic nature of the
step function.  Difficulties with this discontinuity are further amplified when
considering direct expansion of the projector. At finite temperatures close to zero, 
we could  use the analytic Fermi-Dirac function, but this involves the computation of matrix 
exponentials and requires the chemical potential {\em a priori} to high precision.
However, purification methods furnish a recursive, analytic, monotonically increasing and 
highly accurate representation of the step function that does not require prior knowledge of the
chemical potential.  The fundamental idea behind our approach is that this representation 
can be used in a direct variation of the density matrix with respect to a perturbed 
Hamiltonian, where perturbations in $H$ can be carried through at each level 
of purification, either exactly or to finite order \cite{ANiklasson04}.  At finite order, 
this theory provides a framework for the computation of density matrix derivatives in 
the $N$-scaling computation of adiabatic response properties by  perturbed projection 
\cite{VWeber04,VWeber05}. At infinite order, i.e. with an exact expansion, the method 
can be used for efficient quantum embedding of local perturbations \cite{ANiklasson04}.

\subsubsection{Exact expansion (infinite order)}

Assume a perturbation in the Hamiltonian,
\begin{equation}
H = H^{(0)} + H'.
\end{equation}
The recursive expansion of the density matrix
\begin{equation}
F_n(F_{n-1}(\ldots F_1(F_0(H^{(0)}+H'))\ldots )) 
\end{equation}
generates the corresponding perturbed sequence,
\begin{equation}\begin{array}{ll}
{{X}}_n &= X_n^{(0)} + \Delta_n,\\
{{X}}_{n+1} &= F_{n}({{X}}_n),\end{array}
\end{equation}
where $X_n^{(0)}$ is the unperturbed sequence generated from $X_{n+1}^{(0)} = F_{n}(X_n^{(0)})$
with $X_0^{(0)} = H^{(0)}$, and the initial perturbation $\Delta_0 = H'$.
The perturbed orthogonal density matrix is given by
\begin{equation}
P = P^{(0)} + \lim_{n \rightarrow \infty} \Delta_n,
\end{equation}
with the purification differences
\begin{equation}\label{PRT}
\Delta_{n+1} = F_{n}(X_n + \Delta_n) - F_{n}(X_n).
\end{equation}
Combined with the second order trace correcting purification, $F_n$ in Eq.\ (\ref{TC2}), 
we have the following recursive scheme for infinite order, orthogonal density matrix perturbation:
\begin{equation} 
\Delta_1 = - H'/(\varepsilon_{\rm max}-\varepsilon_{\rm min}),
\end{equation}
\begin{equation} \label{DYS_SP2}
\Delta_{n+1} =
\left\{\begin{array}{ll}
\{ X_n^{(0)},\Delta_n\} + \Delta_n^2, & {N_{\rm occ}} \geq N_{\rm e} \\
2\Delta_n - \{ X_n^{(0)},\Delta_n\} - \Delta_n^2, & {N_{\rm occ}} < N_{\rm e} ,
\end{array} \right.
\end{equation}
where we use the anti-commutator notation $\{A,B\}=AB+BA$ and the occupation ${N_{\rm occ}}  = {\rm Tr}(X_n^{(0)})$.
Since the differences $\Delta_n$ change quadratically in each iteration, the
expansion order is in practice infinite at convergence and therefore exact.
For insulators the computational cost scales linearly with the size of the
perturbed region $O(N_{\rm pert.})$ since the recursion only involves terms with the response
factors $\Delta_n$. For a local perturbation the computational cost is therefore
independent of system size, i.e. it scales as $O(1)$ \cite{ANiklasson04}.

The perturbation theory is grand canonical since the expansion of the perturbation is performed
at a fixed chemical potential determined by the unperturbed (or the perturbed) system. For sufficiently 
large perturbations, states may cross the chemical potentials $\mu$. In this case 
$\lim_{n \rightarrow \infty} {\rm Tr}(\Delta_n) \neq 0$ and the system is no longer neutral.

\subsubsection{Finite perturbation expansion}

Assume a perturbation expansion of the Hamiltonian,
\begin{equation}
H = H^{(0)} + \lambda H^{(1)} + \lambda^2 H^{(2)} + \ldots ~ .
\end{equation}
This perturbation generates the corresponding perturbed sequence
\begin{equation}
X_n = X_n^{(0)} + \lambda X^{(1)}_n + \lambda^2 X^{(2)}_n + \ldots ~ ,
\end{equation}
where the separate $m^{\rm th}$ order perturbations $X^{(m)}_n$ can be collected
order by order. Using the second order trace correcting scheme and
keeping terms through order $m$ in $\lambda$ at each iteration,
the following explicit recursive sequence is obtained
for $m = m_{\rm max},m_{\rm max}-1,\ldots ,1,0$:
\begin{equation}\label{second1}
X^{(m)}_{n+1} =
\left\{ \begin{array}{ll}
\displaystyle \sum_{i=0}^{m} X^{(i)}_n X^{(m-i)}_n, &
{N_{\rm occ}} \geq N_{\rm e}\\
\displaystyle 2X^{(m)}_n - \sum_{i=0}^{m} X^{(i)}_n X^{(m-i)}_n,
& {N_{\rm occ}} < N_{\rm e}.\\
\end{array} \right.
\end{equation}
The occupation ${N_{\rm occ}} = {{\rm Tr}}(X^{(0)}_n)$.
The density matrix derivatives are given by
\begin{equation}
\frac{1}{m !}\frac{\partial^m P}{\partial \lambda^m} \bigg|_{\lambda = 0} = P^{(m)}=
\lim_{n\rightarrow\infty} X^{(m)}_{n},
\end{equation}
such that
\begin{equation}
P = P^{(0)} + \lambda P^{(1)} + \lambda^2 P^{(2)} + \ldots ~ .
\end{equation}
These equations provide an explicit and rapidly convergent algorithm for the computation of 
the density matrix response to high order in the expansion parameter \cite{ANiklasson04}.  
In addition, the formalism presented here is remarkably simple and can be easily extended 
to multiple independent perturbations \cite{VWeber04,VWeber05}.

\section{Non-orthogonal Density Matrix Purification and Density Matrix Perturbation Theory}

In a non-orthogonal representation, the Hartree-Fock or Kohn-Sham equations may be posed as the 
generalized matrix eigenvalue problem,
\begin{equation}\label{SES}
{\rm H}\phi_i = \varepsilon_i {\rm S} \phi_i, \quad  \varepsilon_1 \leq \varepsilon_2 \leq \ldots ,
\end{equation}
where the overlap matrix S is a matrix of basis function inner products.

In the following, a non-orthogonal density matrix purification algorithm is developed.  Then,
a non-orthogonal density matrix perturbation theory is introduced that admits simultaneous 
perturbations in both the Hamiltonian and the overlap matrix.  Normal letters (H) are used to 
distinguish the non-orthogonal representation from the corresponding orthogonal representation 
denoted by symbols in italics ($H$). 

\subsection{Non-Orthogonal Purification}

In the non-orthogonal case, the necessary criteria determining the density matrix  are
\begin{equation}\label{CRIT_S}\begin{array}{l}
S{\rm PH-HPS} = 0, \\
{\rm Tr}({\rm PS}) = N_{\rm e}, \\
{\rm P = PSP}, \\
\end{array}
\end{equation}
together with Aufbau filling, i.e. occupying the $N_{\rm e}$ lowest eigenstates.

Following normalization, X$_1 = {\rm F}_0({\rm H})$, purification proceeds as in the orthogonal case, 
but with the minor addition of the metric S to each (previously orthogonal) matrix-matrix multiplication, i.e.
\begin{equation}\label{Metric}
Z = XY ~~ {\rightarrow} ~~ {\rm Z} = {\rm XSY}.
\end{equation}
For the second order trace correction purification, Eq.\ (\ref{TC2}),
\begin{equation}\begin{array}{lll}
F_n(X) &= X^2 ~~ &{\rightarrow} ~~{\rm F}_n({\rm X}) = {\rm XSX},\\
F_n(X) &= 2X - X^2  &{\rightarrow} ~~{\rm F}_n({\rm X}) = {\rm 2X-XSX}.
\end{array}
\end{equation}
This preserves the covariant (or contravariant) form after each purification step \cite{CWhite97}.

The most challenging aspect of non-orthogonal purification is obtaining an initial normalization X$_1$,
which must obey the commutation relation
\begin{equation}\label{INIT}
{\rm SX}_1{\rm H  - HX}_1{\rm S} = 0.
\end{equation}
With this normalization, commutation is automatically preserved as long as
\begin{equation}
{\rm SF}_n({\rm X}_n){\rm H}-{\rm H}{\rm F}_n({\rm X}_n){\rm S} = 
{\rm SX}_n{\rm H}-{\rm H}{\rm X}_n{\rm S}. 
\end{equation}
This is true for all non-orthogonal purification polynomials, which have the form
\begin{equation}
{\rm F}({\rm X}) = \sum_{k=0}^{t} a_k {\rm X}({\rm SX})^k,
\end{equation}
where the $a_k$ are polynomial expansion coefficients.

There are a number of options for initiating the non-orthogonal purification, three of which are
\begin{eqnarray}
\label{INIT1}
\displaystyle {\rm X}_1 = {\rm F}_0({\rm H}) = \alpha(\beta {\rm S}^{-1} - {\rm S}^{-1}{\rm H}{\rm S}^{-1}), \\
\label{INIT2}
\displaystyle {\rm X}_1 = {\rm F}_0({\rm H}) = \alpha({\rm H}^{-1} + \beta {\rm S}^{-1}), \\
\label{INIT3}
\displaystyle {\rm X}_1 = {\rm F}_0({\rm H}) = \alpha({\rm H} - \beta {\rm S})^{-1},
\end{eqnarray}
where $\alpha$ and $\beta$ are chosen to map the eigenvalues into [0,1] in reverse order.
Ideally, an efficient choice of normalization concentrates all unoccupied states near 0, 
and all occupied states near 1. 

The first choice, Eq.\ (\ref{INIT1}), is analogous to the initial guess suggested by Palser and Manolopoulos 
\cite{APalser98} for their non-orthogonal grand canonical purification scheme, and amounts to a shift and 
linear rescaling of eigenvalues.  This linear rescaling can lead to a small renormalized gap in $[0,1]$
for the case of low occupation in the large basis set limit.  The second case, Eq.\ (\ref{INIT2}), is of 
minor interest since it involves calculations of both ${\rm H}^{-1}$ and ${\rm S}^{-1}$. 
The third normalization, Eq.\ (\ref{INIT3}), is the most interesting and useful of the initializations.  
With $\alpha = 1$ and $\beta = \varepsilon_{\rm min} -1$, where $\varepsilon_{\rm min}$
is a lower bound of the eigenvalues $\varepsilon_i$ in Eq.\ (\ref{SES}), this is a
Green's function,
\begin{equation} \label{GNorm}
{\rm G}(\beta) = ({\rm H} -  \beta {\rm S})^{-1},
\end{equation}
which provides the correct normalization.  In this normalization, the unoccupied states
are mapped to $0$ as $1/\varepsilon_i$.  For large basis sets, with a low fractional occupation, 
this amounts to a rescaled band gap on the interval $[0,1]$ that is larger relative to
the band gap given by the linear rescaling. Since the number
of iterations needed to reach convergence scales with the logarithm of the inverse band gap \cite{ANiklasson02A}
the Green's function initialization can be expected to be more efficient in the large
basis set limit.

\subsubsection{Computation and refinement of ${\rm X}_1 = {\rm G}(\beta )$}

The Green's function resolvent ${\rm G}(\beta)$ can be calculated with linear scaling complexity
for sufficiently large and sparse systems using several techniques, such as the Schulz iteration \cite{GSchulz33},
the sparse approximate inverse \cite{MBenzi96,MChallacombe99}, and other methods \cite{TOzaki01}. 
In a self-consistent calculation, where the Hamiltonian is changed in each iteration, or in
a quantum molecular dynamics simulation, where both the overlap and the Hamiltonian is modified, 
we can efficiently update the new initialization 
\begin{equation}
{\rm X}_{1({\rm new})}  = ({\rm H_{new}} -  \beta {\rm S}_{\rm new})^{-1}
\end{equation}
from the previous iteration. If  $X_{1({\rm old})}$ and ${\rm X}_{1({\rm new})}$ are sufficiently close
the following scheme, based on Schulz's method \cite{GSchulz33}, rapidly converges
to the new normalization:
\begin{equation}\begin{array}{ll}
\displaystyle {\rm Y}_0 &= X_{1({\rm old})},\\
\displaystyle {\rm Y}_{n+1} &= 2{\rm Y}_{n} - {\rm Y}_{n}({\rm H}_{\rm new} -  \beta_{\rm new} {\rm S}_{\rm new}){\rm Y}_{n},\\
\displaystyle {\rm X}_{1({\rm new})} &= {\rm G}_{\rm new}(\beta) = \lim_{n \rightarrow \infty} {\rm Y}_{n}.
\end{array}
\end{equation}
In this way, the cost can be reduced by using Schulz's method as an efficient iterative refinement technique. 

\subsubsection{Non-Orthogonal Trace Correcting Purification}

A non-orthogonal second order trace correcting purification scheme is given by
\begin{equation} \label{Pract1S}
{\rm X}_1 = {\rm G}(\beta)  = \left[{\rm H} -  (\varepsilon_{\rm min} -1) {\rm S}\right]^{-1},\\
\end{equation}
\begin{equation}\label{TC2S}
{\rm X}_{n+1} = {\rm F}_n({\rm X}_n) = {\rm X}_n + \sigma_n (I-{\rm X}_n{\rm S}){\rm X}_n,
\end{equation}
where
\begin{equation}\label{Pract2S}
\sigma_n = {\rm sign}[N_{\rm e} - N_{\rm occ}]
\end{equation}
\begin{equation}
N_{\rm occ}={\rm Tr}({\rm S}{\rm X}_n)]
\end{equation}
and
\begin{equation}
{\rm P} = \lim_{n \rightarrow \infty} {\rm X}_n.
\end{equation}
Note that any of the initializations in Eqs.\ (\ref{INIT1})-(\ref{INIT3}) can be used to 
calculate ${\rm X}_1$, realizing  different levels of  performance.

\subsection{Non-Orthogonal Perturbation Theory}\label{NOPRT}

With an efficient normalization scheme in hand, given by Eq.\ (\ref{INIT3}), generalization of the
density matrix perturbation theory to a non-orthogonal formulation follows, constituting
the central result of this paper. At finite order, it provides the framework for computation of
basis set dependent response properties, including magnetic response and geometric energy 
derivatives. The non-orthogonal extension is also useful for density matrix extrapolation
in geometry optimization \cite{ANiklasson05B}.

The non-orthogonal perturbation theory below is developed in the
context of second order trace correcting purification. However, the formalism
can be based also on other purification methods, such as grand canonical 
purification \cite{APalser98,AHolas01,ANiklasson02A}, canonical purification \cite{APalser98},
higher order trace correcting schemes and their hybrids \cite{ANiklasson02A,ANiklasson03,DMazziotti03}, 
implicit purification at finite temperatures \cite{ANiklasson03B}, or matrix sign function 
expansions \cite{GBeylkin99,KNemeth00}.

\subsubsection{Exact expansion (infinite order)}

Assume a perturbation of the Hamitonian and overlap matrix,
\begin{equation}\begin{array}{l}
{\rm H} = {\rm H}^{(0)} + {\rm H'},\\
{\rm S} = {\rm S}^{(0)} + {\rm S'}.
\end{array}
\end{equation}
By analogy with Eq.~(\ref{PRT}), with $X_n$ and $F_n(X_n)$ replaced by
the non-orthogonal sequence ${\rm X}_n$ and purification polynomials ${\rm F}_n({\rm X}_n)$,
we have the non-orthogonal perturbations 
\begin{equation}\label{InitS}
\Delta_1 = {\rm F}_0({\rm H}^{(0)}+{\rm H'})\big|_{\rm S} - {\rm F}_0({\rm H}^{(0)})\big|_{{\rm S}^{(0)}}~~,
\end{equation}
\begin{equation}\label{PRTS}
\Delta_{n+1} = {\rm F}_n({\rm X}_n+\Delta_n)\big|_{\rm S} -  {\rm F}_n({\rm X}_n)\big|_{{\rm S}^{(0)}}~~.
\end{equation}
Here ${\rm F}({\rm X})\big|_{\rm A}$ denotes the non-orthogonal purification with the metrics
adapted to the overlap matrix A.
With ${\rm F}_0({\rm H})$ initiated by a non-orthogonal normalization in 
Eqs.\ (\ref{INIT1})-(\ref{INIT3}) and ${\rm F}_{n}({\rm X}_n)$ by Eq.\ (\ref{TC2S}) we have
\begin{equation}
\Delta_{n+1} = \left\{\begin{array}{ll}
{\rm U}_n, ~~ &{N_{\rm occ}} \geq N_{\rm e} \\
2\Delta_n - {\rm U}_n, ~ &{N_{\rm occ}} < N_{\rm e}, \end{array} \right.
\end{equation}
where
\begin{equation}
{\rm U}_n = \Delta_n{\rm S}({\rm X}_n + \Delta_n) + {\rm X}_n({\rm S}\Delta_n+{\rm S'X}_n).
\end{equation}
The occupation ${N_{\rm occ}} = {\rm Tr}({\rm S}^{(0)}{\rm X}_n)$.
Sine the perturbation theory is based on a perturbed projection, i.e. the difference between the 
purification of the perturbed and unperturbed sequence, the covariant  (or contravariant) 
form of the difference $\Delta_n$ is preserved. This holds true also for the finite 
perturbation expansion described below. At convergence the non-orthogonal perturbed density matrix is
\begin{equation}
{\rm P} = {\rm P}^{(0)} + \lim_{n \rightarrow \infty} \Delta_n.
\end{equation}

\subsubsection{Finite order perturbation}

Assume a perturbation expansion of the Hamiltonian and the overlap matrix, where
\begin{equation}\begin{array}{ll}
{\rm H} &= {\rm H}^{(0)} + \lambda {\rm H}^{(1)} + \lambda^2 {\rm H}^{(2)} + \ldots ~,\\
{\rm S} &= {\rm S}^{(0)} + \lambda {\rm S}^{(1)} + \lambda^2 {\rm S}^{(2)} + \ldots ~.
\end{array}
\end{equation}
These perturbations generates a recursive purification sequence that can be expanded to all orders in $\lambda$,
\begin{equation}
{\rm X}_n = {{\rm X}_n}^{(0)} + \lambda {{\rm X}_n}^{(1)} + \lambda^2 {{\rm X}_n}^{(2)} + \ldots ~.
\end{equation}
The initial expansion of ${\rm X}_1$ can be calculated using any of the normalizations 
given by Eqs.\ (\ref{INIT1})-(\ref{INIT3}).
However, using the Green's function approach in Eq.\ (\ref{INIT3}) makes the
expansion of ${\rm X}_1$ particularly simple. The terms are 
\begin{equation}\label{START}
\begin{array}{ll}
{\rm X}^{(0)}_1 &=  {\rm G}^{(0)} \\
{\rm X}^{(1)}_1 &= -{\rm G}^{(0)} {\rm T}^{(1)} {\rm G}^{(0)}, \\
{\rm X}^{(2)}_1 &= -{\rm G}^{(0)} {\rm T}^{(2)} {\rm G}^{(0)} + {\rm G}^{(0)} {\rm T}^{(1)} {\rm G}^{(0)} {\rm T}^{(1)} {\rm G}^{(0)}, \\
{\rm X}^{(3)}_1 &= -{\rm G}^{(0)} {\rm T}^{(3)} {\rm G}^{(0)} + {\rm G}^{(0)} {\rm T}^{(1)} {\rm G}^{(0)} {\rm T}^{(2)} {\rm G}^{(0)} + \ldots ~,
\end{array}
\end{equation}
where
\begin{equation}
{\rm T}^{(m)} = ({\rm H}^{(m)}-\beta {\rm S}^{(m)}),
\end{equation}
and
\begin{equation}\label{STARTG}
{\rm G}^{(0)} = ({\rm H}^{(0)}-\beta {\rm S}^{(0)})^{-1}.
\end{equation}
This initialization to various order, ${\rm X}^{(m)}_1$, in Eq.\ (\ref{START}) derives from the Dyson series
\begin{equation}
{\rm G} = {\rm G}^{(0)} - {\rm G}^{(0)}\left[{\rm T}^{(1)} + {\rm T}^{(2)} + \ldots \right]{\rm G}^{(0)} + \ldots,
\end{equation}
from which terms in ${\rm G}$ can be collected order by order in $\lambda$. 
A generalization to any order is straightforward.
After the initialization of ${\rm X}^{(m)}_1$ we have for $m = m_{\rm max}, m_{\rm max}-1 , \ldots ,1, 0$:
\begin{equation}\label{PRT_S}
{\rm X}^{(m)}_{n+1} = \left\{\begin{array}{ll}
\displaystyle \sum_{i+j+k=m} {\rm X}^{(i)}_n {\rm S}^{(j)} {\rm X}^{(k)}_n, & {N_{\rm occ}} \geq N_{\rm e} \\
\displaystyle 2{\rm X}^{(m)}_n - \sum_{i+j+k=m} {\rm X}^{(i)}_n {\rm S}^{(j)} {\rm X}^{(k)}_n, & {N_{\rm occ}} < N_{\rm e}.
\end{array} \right.
\end{equation}
The sum is taken over all combinations $(0,1,\ldots,m)$ of $i,j$ and $k$ such that
$i+j+k= m$. The occupation ${N_{\rm occ}} = {\rm Tr}({\rm S}^{(0)} {\rm X}^{(0)}_n)$.
At convergence the density matrix derivatives are given by
\begin{equation}\label{DMD}
\frac{1}{m !}\frac{\partial^m {\rm P}}{\partial \lambda^m} \bigg|_{\lambda = 0} = {\rm P}^{(m)}=
\lim_{n\rightarrow\infty} {\rm X}^{(m)}_{n},
\end{equation}
such that the density matrix perturbation expansion in a non-orthogonal representation is 
\begin{equation}
{\rm P} = {\rm P}^{(0)} + \lambda {\rm P}^{(1)} + \lambda^2 {\rm P}^{(2)} + \ldots ~.
\end{equation}
This method for the calculation of density matrix response, including perturbations in the overlap matrix
for a non-orthogonal representation, composes the central result of this paper.

\section{Example}

\begin{figure}[t]
\resizebox*{3.0in}{!}{\includegraphics[angle=00]{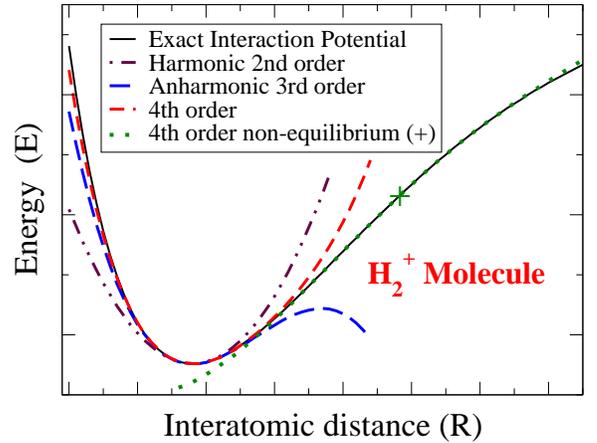}}
\caption{\label{H2}
The analytic expansion of the energy $E(R)$, Eq.\ (\ref{Energy}), as a function of inter-atomic 
distance $R$ for the H$_2^+$ molecule, using the non-orthogonal density matrix perturbation theory up to
$4^{\rm th}$ order.}
\end{figure}

To illustrate the non-orthogonal perturbation theory we have chosen a diatomic
hydrogen ion H$_2^+$ described in a basis set of two hydrogenic 1s-orbitals \cite{PAtkins97}.
The overlap matrix ${\rm S}(R)$ as a function of inter-atomic distance $R$ (in units of Bohr radius $a_0$) 
is given by
\begin{equation}\begin{array}{l}
{\rm S}_{1,1} = {\rm S}_{2,2} = 1\\
{\rm S}_{1,2} = {\rm S}_{2,1} = (1+R+(1/3)R^2)e^{-R}.
\end{array}
\end{equation}
The matrix elements of the Hamiltonian H$(R)$ are
\begin{equation}\begin{array}{l}
{\rm H}_{1,1} = {\rm H}_{2,2} = E_0 -R^{-1}(1-(1+R)e^{-2R}) + \kappa R^{-1}\\
{\rm H}_{1,2} = {\rm H}_{2,1} = (E_0 + \kappa R^{-1}){\rm S}_{1,2}-\kappa{a_0}^{-1}(1+R)e^{-R},
\end{array}
\end{equation}
where $\kappa = e^2/(4\pi \varepsilon_0)$ (set to $1$ in the calculation).
By expanding H and ${\rm S}$ in $r  = (R-R_0)$ around the equilibrium distance $R_0$ (or any other point) we have
\begin{equation}\begin{array}{ll}
{\rm H} &= {\rm H}^{(0)} + r {\rm H}^{(1)} + r^2 {\rm H}^{(2)} + \ldots ~,\\
{\rm S} &= {\rm S}^{(0)} + r {\rm S}^{(1)} + r^2 {\rm S}^{(2)} + \ldots ~,
\end{array}
\end{equation}
where
\begin{equation}\begin{array}{lll}
{\rm H}^{(m)} &= (m!)^{-1}\partial^m{\rm H}/\partial R^m & {\rm at} \quad R = R_0,\\
{\rm S}^{(m)} &= (m!)^{-1}\partial^m{\rm S}/\partial R^m & {\rm at} \quad R = R_0.
\end{array}
\end{equation}
The initial perturbations $\Delta^{(m)}_1$ are given by Eq.\ (\ref{START})
and the recursive expansion is calculated as described in Eq.\ (\ref{PRT_S}).
At convergence the normalized density matrix derivatives ${\rm P}^{(m)}$ are given by Eq.\ (\ref{DMD}).
The expansion of the energy is given by collecting the energy
\begin{equation}\label{Energy}
{E} = {\rm Tr}\left[ ({\rm H}^{(0)} + r {\rm H}^{(1)} + \ldots )({\rm P}^{(0)} + r {\rm P}^{(1)} + \ldots)\right]
\end{equation}
in orders of $r= (R-R_0)$. Figure \ref{H2} shows the interaction potential $E(R)$
as a function of inter-atomic distance in comparison to perturbation expansions up to $4^{\rm th}$ order 
at the equilibrium distance and at $4^{\rm th}$ order at a non-equilibrium inter-atomic distance. 
                                                                                               
\section{Discussion and Conclusions}

In this paper we have shown how density matrix perturbation theory based on recursive purification 
can be generalized to include basis-set dependent perturbations. This makes
it possible, for example, to calculate structural response properties 
using local atomic-centered orbitals within a reduced complexity formalism. 
Some key features of importance are: (1) an orbital-free density matrix formulation,
which avoids the calculation of eigenfunctions and eigenvalues, (2) very high order, monotonically 
increasing analytic approximation of the step function, (3) initial normalization of the Hamiltonian 
to fulfill the non-orthogonal commutation relation, which is preserved after each purification, 
and (4) the ability to collect perturbations recursively, exactly (infinite order) or to any finite 
order, at each level of purification.

A practical generalization of the
Green's function initialization in Eq.\ (\ref{GNorm}) is given by
\begin{equation}
{\rm X}_1 = {\rm G}(z) = z^{-1}\left[({\rm H} - (\varepsilon_{\rm min} - z^{-1})S\right]^{-1},
\end{equation}
which is stable for all $z > 0$. The value of $z$ can be tuned to improve convergence and computational
efficiency by optimizing the size of the band gap of the normalized spectra of ${\rm X}_1$. 
The purification expansion is stable with respect to a complex generalization and the
constant $z$ can be extended to regions of the complex plane, in analogy to 
Green's functions for complex energies. 

If an ill-conditioned non-orthogonal basis set is used we may run into numerical problems if we chose to
transform the generalized eigenvalue problem to an orthogonal representation. 
With the present formulation for non-orthogonal purification and perturbation theory, this congruence transform 
is avoided. Instead it is replaced by the calculation of ${\rm G}(z)$.  However, if the condition number of 
${\rm G}(z)$ is smaller compared to the condition number of ${\rm S}$, the numerical accuracy is improved.
In addition, the back-transform from the orthogonal density matrix representation to the atomic orbital 
representation, which is necessary to calculate the electronic density expressed in the atomic
orbital basis, is avoided within a purely non-orthogonal formalism.

The example for the H$_2^+$ molecule illustrates the extension of the orbital-free density matrix
perturbation theory to non-orthogonal representations. We have also applied the non-orthogonal
method to recalculate the polarizability of molecular clusters with results
identical to previous calculations \cite{VWeber04,VWeber05}. Since 
only matrix-matrix operations are used, the computational cost 
scales linearly with system size for sufficiently large non-metallic systems, 
as was shown previously for the perturbed projection scheme in an orthogonalized representation 
\cite{VWeber04}.  The non-orthogonal density matrix perturbation theory can therefore efficiently 
be applied in calculations of response properties with a perturbation dependent basis set
for large complex systems. 

\section{Acknowledgment}

Discussions with C.\ J. Tymczak and J. Wills are gratefully acknowledged.

\bibliography{mondo_new}

{${\dagger}$ Corresponding author: Anders M.\ N. Niklasson, Email: amn@lanl.gov}

\end{document}